\documentclass[12pt]{spieman}  
\usepackage{amsmath,amsfonts,amssymb}
\usepackage{graphicx}
\usepackage{setspace}
\usepackage{tocloft}

\title{Speckle-illumination spatial frequency domain imaging with a stereo laparoscope for profile-corrected optical property mapping}

\author[a]{Anthony A. Song}
\author[a]{Mason T. Chen}
\author[a]{Taylor L. Bobrow}
\author[a]{Nicholas J. Durr}
\affil[a]{The Johns Hopkins University, Department of Biomedical Engineering, 3400 N. Charles Street, Baltimore, MD, 21218}

\cftpagenumbersoff{figure}
\cftpagenumbersoff{table} 
\begin{document} 
\maketitle

\begin{abstract}\\

\textbf{Significance:}
\noindent
Laparoscopic surgery presents challenges in localizing oncological margins due to poor contrast between healthy and malignant tissues. Optical properties can uniquely identify various tissue types and disease states with high sensitivity and specificity, making it a promising tool for surgical guidance. While spatial frequency domain imaging (SFDI) effectively measures quantitative optical properties, its deployment in laparoscopy is challenging due to the constrained imaging environment. Thus, there is a need for compact structured illumination techniques to enable accurate, quantitative endogenous contrast in minimally invasive surgery.
\\

\textbf{Aim:}
\noindent
We introduce a compact, two-camera laparoscope that incorporates both active stereo depth estimation and speckle-illumination spatial frequency domain imaging (si-SFDI) to map profile-corrected, pixel-level absorption ($\mu_{a}$) and reduced scattering ($\mu^{'}_{s}$) optical properties in images of tissues with complex geometries. 
\\

\textbf{Approach:}
\noindent
We used a multimode fiber-coupled $639nm$ laser illumination to generate high-contrast speckle patterns on the object. These patterns were imaged through a modified commercial stereo laparoscope for optical property estimation via si-SFDI. Compared to the original si-SFDI work, which required $\geq10$ images of randomized speckle patterns for accurate optical property estimations, our approach approximates the DC response using a laser speckle reducer and consequently requires only $2$ images. Additionally, we demonstrate 3D profilometry using active stereo from low-coherence RGB laser flood illumination. Sample topography was then used to correct for measured intensity variations caused by object height and surface angle differences with respect to a calibration phantom. The low-contrast RGB speckle pattern was blurred using a laser speckle reducer to approximate incoherent white-light illumination. We validated profile-corrected si-SFDI against conventional SFDI in phantoms with simple and complex geometries, as well as in a human finger \textit{in-vivo} time-series constriction study.
\\

\textbf{Results:}
\noindent
Laparoscopic si-SFDI optical property measurements agreed with conventional SFDI measurements when measuring flat tissue phantoms, exhibiting an error of $6.4\%$ for absorption and $5.8\%$ for reduced scattering. Profile-correction improved the accuracy for measurements of phantoms with complex geometries, particularly for absorption, where it reduced the error by $23.7\%$. An \textit{in-vivo} finger constriction study further validated laparoscopic si-SFDI, demonstrating an error of $8.2\%$ for absorption and $5.8\%$ for reduced scattering compared to conventional SFDI. Moreover, the observed trends in optical properties due to physiological changes were consistent with previous studies.
\\ 

\textbf{Conclusions:}
\noindent
Our stereo-laparoscopic implementation of si-SFDI provides a simple method to obtain accurate optical property maps through a laparoscope for flat and complex geometries. This has the potential to provide quantitative endogenous contrast for minimally invasive surgical guidance. 
\\
\end{abstract} 

\keywords{Spatial Frequency Domain Imaging, Image-guided surgery, Laparoscopy, Speckle-illumination, Optical properties, Profile correction}

{\noindent \footnotesize\textbf{*}Nicholas J. Durr,  \linkable{ndurr@jhu.edu} }

\begin{spacing}{2}   

\section{Introduction}
\label{sect:intro}  

Laparoscopy is a minimally invasive surgical technique that has become the standard of care for many procedures, such as prostatectomies and appendectomies \cite{ tsui2013minimally}. Compared to traditional open surgery, laparoscopy results in faster recovery times, reduced scarring, and lower medical costs \cite{agha2003does}. However, laparoscopy presents challenges in visualization due to limited field of view (FOV) and poor contrast, which can make it difficult to accurately identify critical anatomical structures, assess tissue perfusion, and determine cancerous tissue margins\cite{boppart1999optical}. These limitations can lead to subjective decision-making reliant on the surgeon’s experience \cite{de2004effect}, leading to wide variability in clinical outcomes \cite{ignjatovic2009optical}. This has spurred the development of optical technologies for surgical guidance, including fluorescence contrast \cite{franz2021tumor, barth2023nerve} and endogenous contrast \cite{baltussen2019hyperspectral, nguyen2013novel, curran2010reflectance}. However, fluorescence imaging using an exogenous dye requires pre- or peri-operative administration of an optical contrast agent, which increases intraoperative workflow complexity. Additionally, many of these techniques are qualitative, leading to user subjectivity, which limits their practical clinical value and generalization to downstream tasks employing artificial intelligence. Recent research studies have demonstrated that tissue optical properties can be used to identify tissue types with high sensitivity and specificity \cite{langhout2018nerve, curran2010reflectance}, underscoring the potential impact of laparoscopic techniques enabling quantitative mapping of optical properties.
\\
\noindent
\\
Spatial frequency domain imaging (SFDI) is a low-cost, non-contact, wide-field, rapid, technique for mapping tissue optical properties \cite{cuccia2009quantitation, gioux2019spatial}. Conventional SFDI involves projecting a 2D sinusoidal pattern of light onto a medium and imaging the back reflectance using a standard camera. A turbid medium scatters the projected light, preferentially attenuating high spatial frequencies. Additionally, absorption by tissue chromophores primarily attenuates the low spatial frequencies of the illumination pattern. The acquired diffuse reflectance images at high and low spatial frequencies are used to decouple and estimate absorption and reduced scattering coefficients via lookup tables (LUTs) generated from Monte Carlo light transport simulations. By acquiring multiple measurements with unique illumination wavelengths and spatial frequencies measurements, the concentrations of dominant tissue chromophores, such as oxy- and deoxy-hemoglobin, can be mapped. Clinically, SFDI has been successfully translated into surgical interventions, such as measuring ischemic onset in skin flaps \cite{gioux2011first, pharaon2010early}, kidney \cite{nadeau2013quantitative}, bowel \cite{gioux2011first}, and liver \cite{gioux2011first}. Additionally, SFDI has shown promise in oncologic margin detection for glioma \cite{sibai2015quantitative}, esophageal cancer \cite{sweer2019wide}, and colon cancer \cite{nandy2018label}.
\\
\noindent
\\
Incorporating conventional SFDI into minimally invasive surgery has been challenged by slow measurement acquisition rates, complex tissue geometries, and compatibility of the projection optics with rigid laparoscopes. Conventional SFDI requires at least six images per wavelength (three distinct spatial phases at two spatial frequencies) to estimate a single optical property map, limiting the acquisition rate to approximately $5$ Hz. To accelerate SFDI towards video-rate imaging, a technique called single snapshot optical properties (SSOP) \cite{angelo2017real} was introduced, which reconstructs an optical property map from a single structured light image. SSOP performs demodulation in the Fourier domain, separating the DC (planar) and AC (spatially modulated) components from a single-phase AC image. To improve the accuracy of optical property estimation in 3D samples, Gioux \textit{et al.} \cite{gioux2009three} used phase profilometry to estimate the 3D sample profile, enabling correction for irradiance variations associated with height and surface angle. There have been numerous approaches to condensing the bulky SFDI projector into an endoscopic form factor. Angelo \textit{et al.} \cite{angelo2017real} introduced real-time endoscopic SSOP by coupling light through a fixed grating into one of the laparoscope ports. Although this endoscopic SSOP approach produces video-rate optical property measurements, it suffers from a number of weaknesses: (1) undesirable artifacts and reduced accuracy due to Fourier filtering of a single image; (2) its illumination strategy occupies one of the laparoscope ports, precluding stereo imaging; and (3) incompatibility with the modern sensor-at-the-tip design of stereo laparoscopes. More recently, Crowley \textit{et al.} \cite{crowley2024ultra} developed an "ultra-miniature" SFDI system that produces fringe patterns interferometrically through a fiber array, making it sufficiently small for a standard colonoscope. However, the fiber array produces non-ideal illumination patterns with poor modulation depth. These patterns are susceptible to noise and cross-core coupling, leading to the degradation of optical property estimation. There is also no way of estimating depth and surface shape in this approach, which is crucial for the accurate estimation of optical properties in real-world applications. 
\\
\noindent
\\
To overcome some of the limitations of previous endoscopic SFDI efforts, we build upon our previous work on speckle-illumination SFDI (si-SFDI) \cite{chen2021speckle}. We introduce a two-camera laparoscopic si-SFDI system coupled with an active stereo profile estimation scheme to generate surface profile-corrected optical property maps. In this study, we estimate optical properties using only two images: one speckle image (AC) and one speckle-reduced image (DC). This approach improves upon previous si-SFDI work, which required $\geq10$ images to approximate the DC image and accurately sample the low spatial frequency MTF response. We propose an alternative technique that approximates the DC image using a laser speckle reducer (LSR). To replace the bulky projector system, which previously required a free space laser expanded to illuminate a diffuser, we introduce multimode fiber-based speckle illumination, which allows compact speckle pattern generation over a wide FOV and large depth of field (DOF). In addition to optical property measurements, laser speckle illumination provides synthetic texture for stereo matching, enabling high-fidelity 3D profile estimations of low-contrast samples \cite{khan2018single}. These 3D reconstructions are then used for downstream SFDI height- and angle-dependent intensity corrections, enabling accurate optical property measurements of non-flat surfaces. We employ two fiber illumination conditions: a high coherence red laser (HC-R) to generate high-contrast speckle patterns for optical property estimation, and a low coherence RGB laser (LC-RGB) to produce low contrast speckle patterns for stereo 3D reconstruction. The LC-RGB source is additionally speckle-reduced to allow speckle-free conventional white light imaging. Overall, this study validates stereo laparoscopic si-SFDI's ability to accurately and precisely measure profile-corrected optical property maps for tissue-mimicking phantoms arranged with complex geometries and quantitative physiological changes of an \textit{in-vivo} sample, paving the way for optical imaging-guided minimally invasive surgery using quantitative endogenous contrast.

\section{Principles of si-SFDI}

Previous work demonstrated that optical properties can be mapped using structured light from laser speckle illumination instead of sinusoidal illumination \cite{chen2021speckle}. Whereas conventional SFDI uses sinusoidal illumination to sample discrete spatial frequencies of a tissue’s modulation transfer function (MTF), si-SFDI uses speckle illumination to sample a wide range of spatial frequency bands of the MTF. si-SFDI relies on the consistency of the power spectral density (PSD), which requires that the integral of the PSD should remain constant for any randomization of the speckle pattern. For an unknown turbid sample, changes in the measured PSD relative to a calibration phantom reflect differences in sample's optical properties. For samples with the same absorption but decreased reduced scattering coefficients, incident photons will diffuse deeper into the media and emerge farther from their entry point. This wider spread of photons blurs the speckle pattern, attenuating higher spatial frequencies and narrowing the PSD. Conversely, in samples with constant scattering but higher absorption, photons are absorbed before they diffuse far from their entry point. This reduces the speckle blur, attenuating lower spatial frequencies and broadening the PSD \cite{jain2019measuring}.
\\
\noindent
\\
Compared to laser speckle contrast imaging, which measures the blur of \textit{subjective} speckle patterns, si-SFDI relies on measuring the blur of the \textit{objective} speckle patterns. Subjective speckle patterns are produced from coherent light interacting with the object and interfering at the detection plane \cite{goodman2007speckle}. Objective speckle, on the other hand, results from interference of the coherent source within a passive optical element that imparts spatially-varying random phase shifts \cite{bonin1989simple}, and is independent of the sample properties and detector position or optics. The previous si-SFDI study produced objective speckle patterns by sending coherent illumination through a diffuser. In this study, we produce objective speckle patterns through modal interference of coherent light in a multimode optical fiber \cite{imai1980speckle, hu2020does}. The mean objective speckle diameter $d_{o}$ at the object plane is inversely proportional to the fiber core radius, $a$:  

\begin{equation}
\label{eq:obj_speckle}
d_{o} = \frac{2 \lambda z }{ \pi a } ,
\end{equation}

\noindent
where $\lambda$ is the wavelength and $z$ is the distance from the exit face of the fiber to the object \cite{takai1985statistical, goodman2007speckle}. In contrast, subjective speckle is pose-dependent and primarily a function of the detection optics \cite{boas2010laser, goodman2007speckle}. The minimum subjective speckle diameter $d_{s}$ can be expressed as: 

\begin{equation}
\label{eq:subj_speckle}
d_{s} = 2.44 \lambda (1+m) F_{\#} ,
\end{equation}

\noindent
where $F_{\#}$ and $m$ represent the detection system’s f-number and magnification, respectively. To extract optical properties from coherently-illuminated samples, the size of $d_{o}$ should be selected to sample the MTF at spatial frequencies that are sensitive to changes in reduced scattering coefficient (e.g. $0.1-0.2mm^{-1}$). Moreover, the average subjective speckle diameter should be sufficiently different than the average objective speckle diameter to allow the removal of subjective speckle in acquired images by spatial filtering. 
\\
\noindent
\\
To generate optical property maps using si-SFDI, we calculate a PSD for each image pixel based on the radially averaged autocorrelation function (ACF) of a sliding window. The size of this window is selected to match the desired sampling spatial frequency. The ACF function $a_{v}$ is estimated using the Wiener-Khinchin theorem \cite{goodman2015statistical} to reduce computational cost: 

\begin{equation}
\label{eq:ACF}
a_{v}(x,y) = F^{-1}[V^{*}(k_{x}, k_{y}) \cdot V(k_{x}, k_{y})],
\end{equation}

\noindent
where, $V$ and $V^{*}$ represent the Fourier transform of the sliding window $v(x,y)$ and its conjugate. The PSD, $S_{v}(k_{r})$, is calculated by taking the magnitude of the Fourier transform of the radially averaged $a_{v}(x,y)$:

\begin{equation}
\label{eq:PSD}
S_{v}(k_{r}) = |H(k_{r})|^{2} \cdot S_{U}(k_{r}),
\end{equation}

\noindent
where, $H(k_{r})$ represents the sample response, and $S_{U}(k_{r})$ represents the system and speckle pattern response. We record the speckle-illumination response of a reference phantom with known optical properties under laser speckle illumination. This allows us to scale the ratio between the PSDs of the sample and reference phantom using the reference model response predicted by Monte Carlo simulations. In this way, we characterize the sample response $H(k_{r})$ due to scattering and absorption without knowing the input pattern $u(x,y)$. We define the AC response $M_{AC}$ as the sum of the frequency responses corresponding to the PSD curve centered around $0.2mm^{-1}$ and $0.4mm^{-1}$, representing the spatial frequency response from $0.1mm^{-1}$ to $0.5mm^{-1}$. The DC response $M_{DC}$ is the remitted signal under planar illumination. Unlike our previous work \cite{chen2021speckle}, which approximated $M_{DC}$ by averaging $10$ random speckle images produced by moving a diffuser to $10$ unique positions, this study uses a laser speckle reducer (LSR) attached to the illumination fiber. The LSR produces multiple speckle patterns through fiber vibration, which are averaged within a single camera exposure, thus approximating $M_{DC}$ with just one image. Using this approximation, we subtract $M_{DC}$ from the speckle image to ensure the calculated ACF has a zero mean.
\\
\noindent
\\
Following the approach outlined in Cuccia \textit{et al.} \cite{cuccia2009quantitation}, the AC and DC diffuse reflectance ($R_{d}$) of a sample are calculated using $R_{d}=\frac{M}{M_{ref}}\cdot R_{d,ref,pred}$, where $M_{ref}$ is the response parameter of the reference phantom, and $R_{d,ref,pred}$ represents the reflectance value predicted by White Monte Carlo models \cite{kienle1996determination}. We use a pre-computed LUT to determine $(\mu_{a}, \mu^{'}_{s})$ from $(R_{d,DC}, R_{d,AC})$ using linear interpolation. The LUT was generated by performing Monte Carlo simulations of $R_{d,AC}$ at $0.1mm^{-1}$ increments from $0.1mm^{-1}$ to $0.5mm^{-1}$ and summing them together \cite{erickson2010lookup}. 

\section{Stereo Measurements for Profile-corrected Optical Property Mapping}

In order to accurately measure $R_{d}$, it is important to correct for intensity variation due to sample height and curvature. We perform profile-correction for non-planar sample surfaces and changes in height, by modeling light reflecting from a diffuse surface with a Lambertian model and approximating the source vector as being coincident with the camera vector. In this model, the illumination intensity varies as $cos(\theta)/z^{2}$ \cite{nayar1989determining}, where $\theta$ is the angle between the sample surface normal and the optical axis of the laparoscope. For example, as an object's distance from the laparoscope tip increases, the image will appear darker, and as $\theta$ increases, it also becomes darker. 

\subsection{Active stereo profile estimation} \label{ssec:3.1}

Recovering depth information from a pair of stereoscopic images requires the identification of matching disparity features between images. In passive stereo imaging, which uses white light flood illumination, extracting stereoscopic depth information from objects with subtle features is challenging due to insufficient disparity candidates \cite{wei2022stereo}. To address this challenge, we adopted an active stereo approach by projecting a unique speckle pattern onto the object. This technique introduces synthetic texture, providing unique and identifiable features that solve the correspondence problem between stereo images. Previous work has shown that structured light imaging improves stereo matching efficiency and improves profile reconstruction accuracy \cite{pribanic2012stereo}. Furthermore, compared to monochromatic structured illumination, RGB structured illumination provides additional color information, enhancing structural contrast and resolving ambiguities caused by intensity-based stereo matching \cite{chen1997range, zhang2002rapid}. Therefore, we projected RGB laser speckle onto the sample to improve pixel-level estimation of height and surface normals. 
\\
\noindent
\\
Stereo camera calibration was performed in MATLAB (R2022a, MathWorks) using $69$ image pairs of a checkerboard pattern ($3\times4$, $2.5mm$ sized squares) at various positions. These positions ranged from $4.8cm$ to $10.8cm$ in height, with surface normal angles up to $40$ degrees relative to the optical axis of the laparoscope. A disparity map between the two active stereo images was generated using a fast coarse-to-fine active stereo matching algorithm \cite{fu2019fast}. The two active stereo images were flat-field corrected with a Gaussian smoothing standard deviation of $\sigma=55$. A 3D point cloud was generated from the disparity map using a reprojection matrix obtained through stereo calibration. Noisy data points were removed through a series of steps: downsampling the 3D point cloud in XY using $2\times2$ spatial averaging, applying Golay filter smoothing with a polynomial order of $1$ and frame length of $31$, and fitting the result to a smooth surface using the \textit{gridfit} method in MATLAB \cite{jderrico2016}. Surface normals were estimated by fitting a local plane using $k=6$ neighboring points and obtaining the normal vector (\textit{pcnormals} in MATLAB).

\subsection{Height correction}\label{ssec:3.2}

To address variations in intensity and speckle size in the object space caused by changes in the object's distance from the laparoscope tip, we implemented a calibration-based profile-correction approach using a reference phantom with known optical properties \cite{gioux2009three, angelo2017real}. 
\\
\noindent
\\
To generate the height-correction model, we acquired laparoscopic si-SFDI and active stereo images of the reference phantom at five different distances, ranging from $5.8cm$ to $9.8cm$ in $1cm$ increments, which represent the typical range of object distances in laparoscopic surgery. The window size was adjusted to ensure that the sliding window samples the desired spatial frequency resolution in the object space at different distances. The spatial frequency resolution defines the interval between consecutive values on the PSD's x-axis. For example, as the working distance increases, the sliding window size decreases, and vice versa. For each distance measurement, we computed the modulation ($M_{AC,ref,z=[5.8, 9.8]}$ and $M_{DC,ref,z=[5.8, 9.8]}$) using si-SFDI. These measurements were fit using a piecewise cubic hermite interpolating polynomial, yielding functions that relate modulation and distance. The $M_{AC,ref,z}$ function primarily characterizes high spatial frequency attenuation (most sensitive to scattering) with distance, while the $M_{DC,ref,z}$ function primarily characterizes low spatial frequency (most sensitive to absorption) attenuation with distance.
\\
\noindent
\\
To generate height-corrected optical property measurements for a test sample, we first acquired a pair of stereo images to generate a sample height map based on the active stereo approach mentioned above. We used the sample height map to determine the appropriate sliding window size for each pixel, ensuring consistent spatial frequency resolution across the height-varying sample. The acquired speckle and speckle-reduced images were then used to compute the uncorrected sample modulation ($M_{AC,sam}$ and $M_{DC,sam}$). We interpolated the $M_{AC,ref,z}$ and $M_{DC,ref,z}$ maps using the sample height map to apply the height-correction model to the new object. This resulted in height-corrected virtual calibration phantoms $M_{AC,ref-height-corr}$ and $M_{DC,ref-height-corr}$. The height-corrected diffuse reflectance of the sample $R_{d,sam,height-corr}$ was calculated as the Monte Carlo predicted reflectance value $R_{d,ref,pred}$, corrected for height effects by multiplying by the ratio of the $M_{AC,sam}$, and the $M_{AC,ref-height-corr}$:

\begin{equation}
\label{eq:height_corr}
R_{d, sam, height-corr} = \frac{M_{AC, sam}}{M_{AC, ref-height-corr}} \cdot R_{d, ref, pred}
\end{equation}

\subsection{Angle-dependent intensity correction} \label{ssec:3.3}

We corrected for illumination variation on object surfaces that are not perpendicular to the illumination axis and assumed Lambertian reflectance and that the illumination vector was the same for all points in the FOV. For this simple model, we assumed that the illumination vector was parallel to the collection axis. 
\\
\noindent
\\
We used the surface normals acquired from the smoothed point cloud to compute the polar angle $\theta$, which is the angle between the surface normal and the illumination vector, for every pixel in the FOV. This angular-correction model was applied as a cosine term in the denominator of $R_{d,sam,height-corr}$ to obtain the height- and angle-corrected diffuse reflectance of the sample\cite{gioux2009three}, denoted as $R_{d,sam,angle-height-corr}$.

\begin{equation}
\label{eq:ang_corr}
R_{d,sam,angle-height-corr} = R_{d,sam,height-corr} \cdot \frac{1}{cos(\theta)}
\end{equation}
\\
\noindent
This angular-correction model increases effective $R_{d}$ as the angle $\theta$ increases.

\section{si-SFDI Laparoscope System}

\begin{figure}
\begin{center}
\begin{tabular}{c}
\includegraphics[width=1\linewidth]{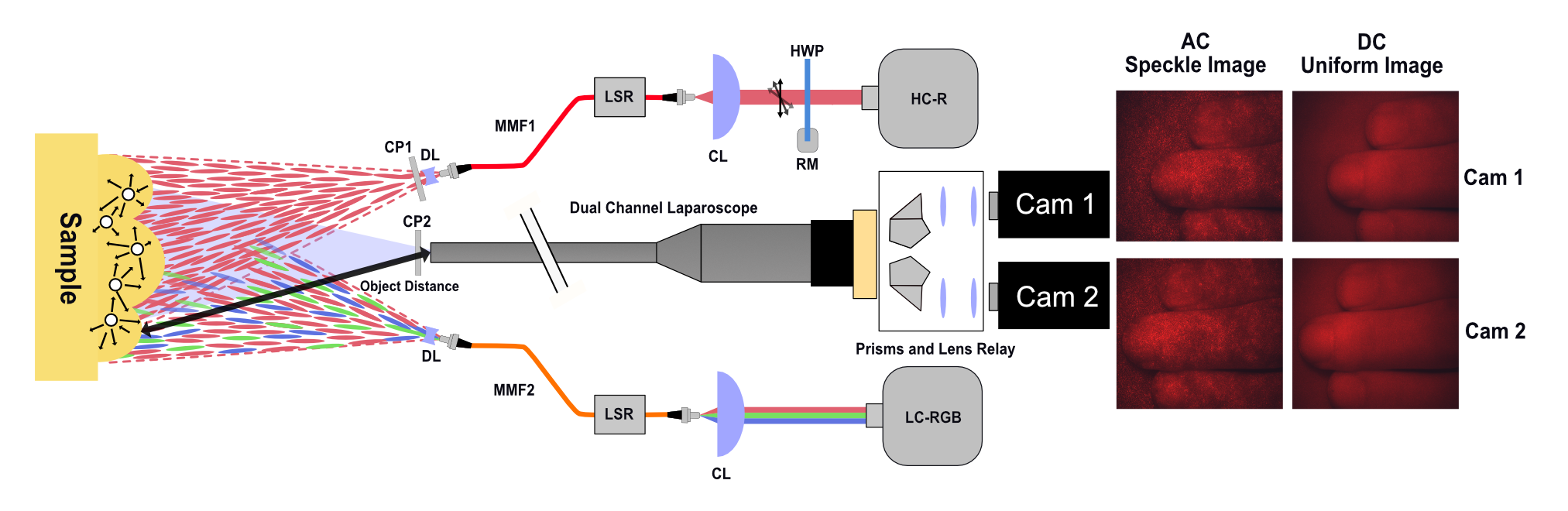}
\end{tabular}
\end{center}
\caption 
{ \label{fig:figure1} \textbf{Laser speckle stereo laparoscope imaging schematics:} A high coherence red laser (HC-R) and low coherence red/green/blue lasers (LC-RGB) source are coupled into multimode fibers (MMF1/2) with coupling lenses (CLs) to illuminate the full laparoscopic field of view. The HC-R light is cross-polarized (CP1/2) with the imaging channel to reduce specular reflection and isolate diffuse reflectance for si-SFDI optical property estimation. A Laser speckle reducer (LSR) is toggled to alternate between high-coherence speckle illumination (AC Speckle Image) and uniform illumination (DC Uniform Image). The LC-RGB illumination provides synthetic texture that improves stereo matching and the resulting profile estimation. Additionally, LC-RGB can be speckle-reduced using a LSR to provide conventional white light images.}
\end{figure} 

Figure~\ref{fig:figure1} shows the experimental setup. A rigid dual camera stereo laparoscope (Intuitive Surgical Scholly $311464-05$ Da Vinci Robotic $12$MM Endoscope $0^{\circ}$ Tip) with relay optics (Intuitive Surgical $370538-04$ Da Vinci Wide Angle Camera Adapter) imaged two views of the object to two $8$-bit RGB CMOS color cameras (GS3-U3-41C6C-C $1$" FLIR Grasshopper $3$ High Performance USB $3.0$ Color Camera) with a stereo baseline of $5mm$. The acquired images were cropped to the sensors’ central $784 \times 960$ pixels to remove obscured pixels.This region of interest (ROI) has an angular FOV (AFOV) of $33 \times 39$ degrees, resulting in a $4.6cm \times 5.6cm$ FOV and $58\mu m$ object pixel size at a distance of $7.8cm$.
\\
\noindent
\\
To generate high contrast red laser objective speckles for measuring optical properties, we used a $639nm$ diode-pumped solid-state single longitudinal mode laser (HC-R; CNI Laser, Changchun, CN, MSL-FN-$639$) with a coherence length greater than $40$ meters. This light was coupled into a $0.50$ NA, $600\mu m$ core step-index multimode fiber optic cable (MMF1; Thorlabs, M53L01) using an aspheric condenser lens (CL; Thorlabs, ACL25416U). We can select a fiber core diameter that produces the ideal speckle size on the object plane. Importantly, $d_{o}$ increases with distance (Eq. (\ref{eq:obj_speckle})). Consequently, when a sample is measured at a greater distance than the reference phantom, the larger projected $d_{o}$ results in a distance-dependent shift towards lower spatial frequencies in the PSD. However, if the illumination and detection AFOVs are equal, the 'apparent' speckle grain size on the image sensor remains constant regardless of distance, and therefore, the pattern's carrier frequencies do not change on the sensor. We can then account for distance-dependent spatial frequency shifts by first calibrating over several known distances and measuring the sample height, as described in Section~\ref{ssec:3.2}.
\\
\noindent
\\
To ensure the AFOVs of the illumination and detection were equal, and to expand the objective speckle size, the light was spread by an endoscopic diverging lens (DL; Olympus, WTLGLA-187W), which was mounted on the fiber tip using a custom 3D-printed mount. We selected an illumination fiber diameter that produced an average objective speckle grain size of $5$ pixels across all distances. This size satisfies the Nyquist criterion for the red channel of our RGB Bayer pattern (GBRG) image sensor.
\\
\noindent
\\
At a $7.8cm$ calibration distance, we used an $85 \times 85$-pixel sliding window to generate the PSD, yielding a spatial frequency resolution of $0.2mm^{-1}$. As mentioned in Section~\ref{ssec:3.2}, the sliding window size should change as a function of distance to ensure consistent spatial frequency resolution of the PSD. We defined the AC response using the first two non-zero spatial frequencies (centered at $0.2mm^{-1}$ and $0.4mm^{-1}$). This approach incorporates the frequency commonly used in SFDI for biological tissue ($0.2mm^{-1}$) while improving stability and reducing noise influence by including the higher frequency component. We used a sliding window step size of $5$ to balance computational speed and spatial resolution of the optical property maps. Smaller steps may improve resolution for highly heterogeneous samples but increase processing time.
\\
\noindent
\\
To tune the output power of the laser, a half-wave plate (HWP) was mounted to a direct drive rotation mount (RM) that controls the beam's polarization angle. At a distance of $7.8cm$, we measured a power density of $2.05 mW/cm^{2}$ for the $639nm$ illumination using a standard photodiode power sensor (Thorlabs, S121C/PM100D). This is well below the OSHA safety limit for Class I lasers, which specifies a maximum permissible exposure (MPE) for skin at $639nm$ of $200 mW/cm^{2}$ for $>10$-second exposure. Additionally, in order to isolate scattered light from the sample and remove specular reflections, we mounted crossed polarizers at the speckle illumination fiber output (CP1) and the laparoscope imaging input (CP2) using two $\varnothing 1/2"$ Linear Polarizers (Thorlabs, LPVISE050-A). The illumination fiber output was mounted at a $15$-degree angle on the side of the laparoscope imaging input to balance proximity to the optical axis with full field-of-view illumination, given the components at the tip.
\\
\noindent
\\
To produce a speckle-reduced, planar illumination (DC) image, a custom-made laser speckle reducer (LSR) was attached to MMF1. The LSR utilizes a vibration motor (UXCELL DC $6$V $3000$RPM Arch Vibrating Wheel) operating at approximately $2000$ RPM. This process causes multiple random speckle patterns within the exposure time of the image capture, approximating a planar illumination (DC) image. As mentioned in Section~\ref{ssec:3.1}, to improve stereo matching through synthetic texture projection, low contrast RGB laser speckle was generated by coupling a low coherence RGB laser (LC-RGB) module (Micro RGB Laser Module, Opt Lasers) into a separate $0.50$ NA, $400\mu m$ core, step-index multimode fiber optic light guide (MMF2). Similar to MMF1, a diverging lens was used to illuminate the full camera FOV and expand the objective speckle size. The LC-RGB module consists of red ($638nm +5/-3nm$), green ($525nm +10/-5nm$), and blue ($450nm \pm10nm$) laser diodes. At a distance of $7.8 cm$, the power density for each RGB laser diode (red: $0.77 mW/cm^{2}$, green: $0.62 mW/cm^{2}$, blue: $0.95 mW/cm^{2}$) was well below the MPE for skin specified by OSHA for Class I lasers at their respective wavelengths. Furthermore, the low contrast RGB speckle was speckle-reduced using the LSR to approximate incoherent white light illumination.
\\
\noindent
\\
To measure profile-corrected optical properties for each sample, we acquired one speckle (AC) image and one speckle-reduced (DC) image, each with a $30ms$ exposure time and two active stereo RGB images, each at a $90ms$ exposure time. For si-SFDI optical property estimation, we used the red channel of AC and DC images, while the RGB images were used for active stereo surface profile estimation. The estimated subjective speckle grain size from Eq. (\ref{eq:subj_speckle}) is $0.92$ pixels \cite{boas2010laser}. This is consistent with experimental images, where subjective speckle grains are approximately $1$ pixel in diameter. To reduce the contribution of the $\approx1$ pixel subjective speckle while maintaining the optical property information encoded in the blurred $\approx5$ pixel objective speckle, we applied a Gaussian filter with a kernel size of $3$ pixels to the raw image. All measurements were taken in a dark room to reduce the influence of ambient light and simulate a low-light, minimally invasive surgical environment. 

\section{Results}
\label{sect:results}  

\subsection{Optical properties measurements of flat tissue-like phantoms} \label{ssec:5.1}

\begin{figure}
\begin{center}
\begin{tabular}{c}
\includegraphics[width=1\linewidth]{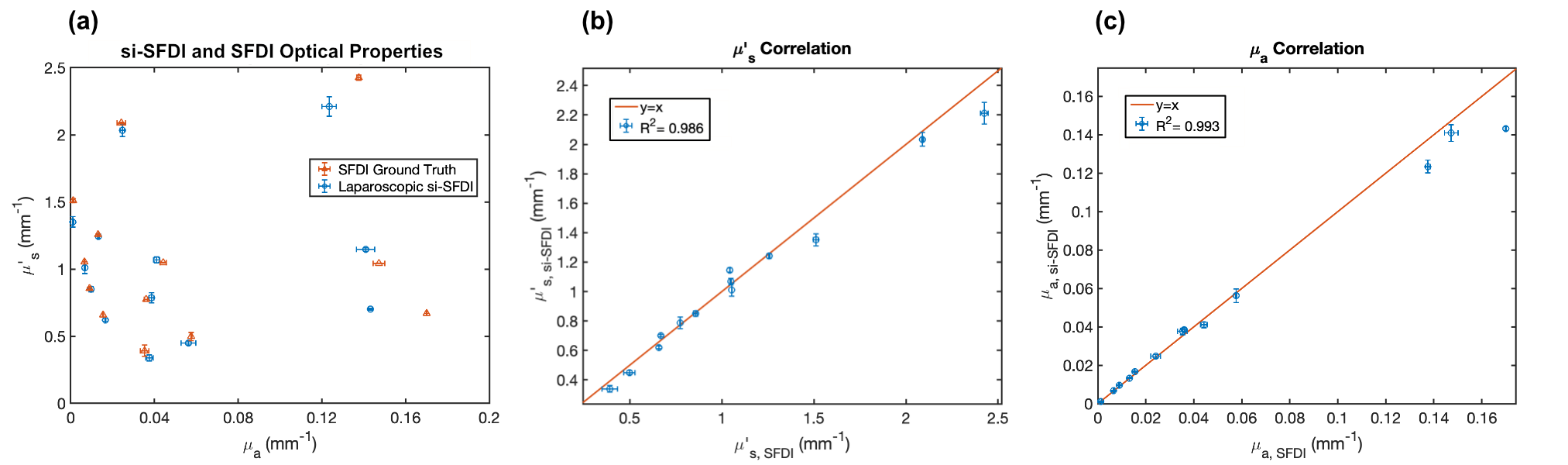}
\end{tabular}
\end{center}
\caption 
{ \label{fig:figure2}\textbf{Comparison of laparoscopic si-SFDI (blue circles) and ground truth SFDI (orange triangles) measured optical properties of 13 homogenous tissue-mimicking phantoms. (a)} si-SFDI measurement versus SFDI ground truth. si-SFDI measurements (y-axis) of \textbf{(b)} reduced scattering and \textbf{(c)} absorption versus SFDI ground truth (x-axis). The coefficient of determination $R^{2}$ is calculated based on the $y=x$ line representing ground truth. Markers indicate mean values and error bars indicate standard deviation.}
\end{figure} 

We first validated our laparoscopic si-SFDI system using $13$ homogeneous tissue-mimicking phantoms with unique combinations of optical properties. The phantoms were fabricated by mixing polydimethylsiloxane (PDMS) with titanium dioxide (TiO$_{2}$) and India Ink for scattering and absorption agents, respectively \cite{greening2014characterization}. We compared optical properties for each phantom estimated by laparoscopic si-SFDI at $639nm$ and conventional three-phase, $0mm^{-1}$ and $0.2mm^{-1}$ spatial frequency SFDI at $659nm$ (Modulim Reflect RS). The calibration phantom used in all of the experiments was a large ($21.5 \times 21.5 \times 2.5cm$), flat tissue-mimicking phantom with known optical properties ($\mu_{a}= 0.024mm^{-1}$, $\mu^{'}_{s}= 0.99mm^{-1}$; $639nm$). To account for the wavelength mismatch between si-SFDI ($639nm$) and SFDI ($659nm$), we corrected the measured sample optical properties using the predicted optical properties of India ink and TiO$_{2}$ at their respective wavelengths. To correct absorption, we used the ratio of India ink's extinction coefficients ($\epsilon$) at the corresponding wavelengths of $639nm$ and $659nm$ \cite{di2010use}, which resulted in $\mu_{a,si-SFDI}=1.0361 \mu_{a, SFDI}$. To correct for reduced scattering, we used a Rayleigh scattering relationship ($\mu^{'}_{s} \propto \lambda^{-4}$), which resulted in $\mu^{'}_{s, si-SFDI}=1.131 \mu^{'}_{s, SFDI}$. We compared the mean and standard deviation for central ROIs between a $100 \times 100$-pixel laparoscopic si-SFDI ROI and a ground truth $150 \times 150$-pixel SFDI ROI. Due to the different imaging scales of si-SFDI and SFDI, we adjusted the sizes of the ROIs to ensure we sampled comparable proportions of the phantom's area across both modalities. Figure~\ref{fig:figure2} shows scatter plots of the optical properties estimated from tissue-mimicking phantoms using laparoscopic si-SFDI with one AC image and one DC image. We observed a mean absolute error of $6.4\%$ for absorption and $5.8\%$ for reduced scattering. We found that laparoscopic si-SFDI predicted optical properties fit a $y=x$ ground truth line with a coefficient of determination $R^{2}$ value of $0.986$ for reduced scattering and $0.993$ for absorption, as shown in Figures~\ref{fig:figure2}(b) and ~\ref{fig:figure2}(c), respectively.

\subsection{Height and angle estimation using active stereo} \label{ssec:5.2}

\begin{figure}
\begin{center}
\begin{tabular}{c}
\includegraphics[width=1\linewidth]{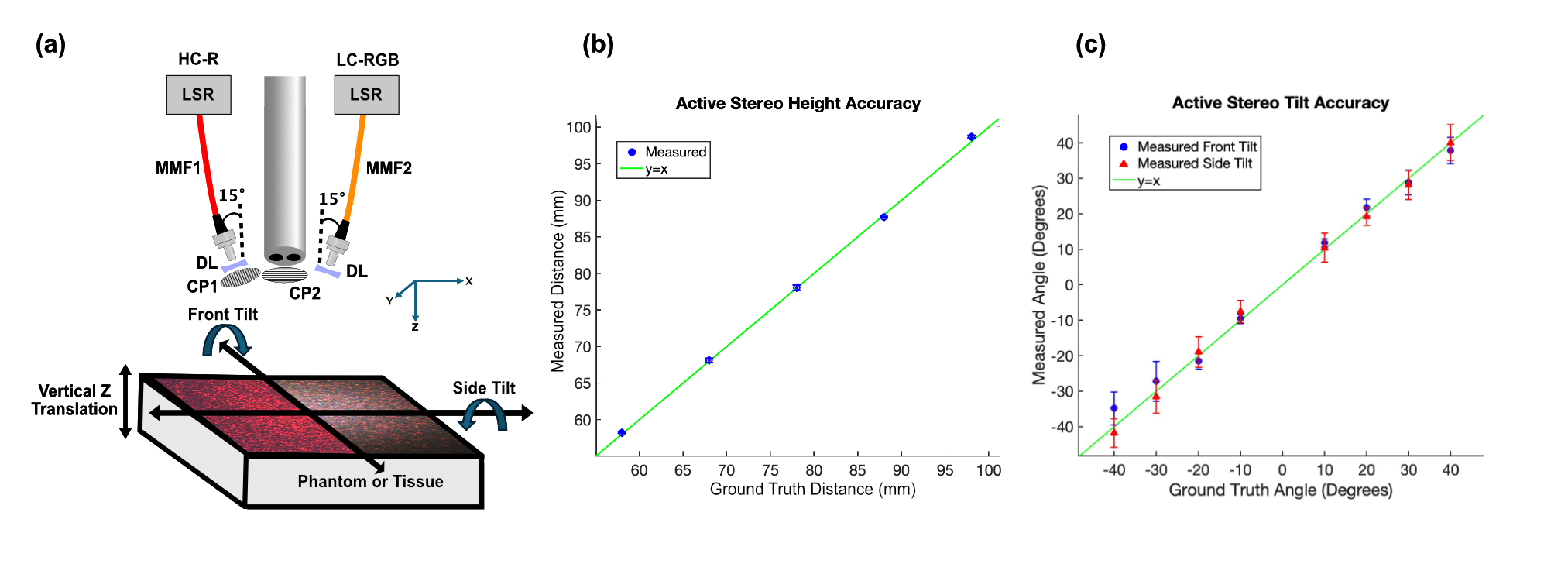}
\end{tabular}
\end{center}
\caption 
{ \label{fig:figure3}\textbf{Sample geometry estimation setup and results: (a)} Close-up schematic of the speckle illumination conditions. The high coherence red laser (HC-R) produces high contrast red laser speckle illumination for optical property estimation and is delivered through the multimode fiber (MMF1). MMF1 illumination is cross-polarized (CP1/2) with the imaging channel to isolate diffuse reflectance. The low coherence RGB lasers (LC-RGB) produce low contrast RGB laser speckle illumination for enhanced stereo matching and white light imaging, and are delivered through MMF2. Light through MMF1 and MMF2 are positioned in the XZ plane at a $15$-degree angle relative to the imaging optical axis. Diverging lenses (DLs) are used to expand the illumination, and laser speckle reducers (LSRs) are used to provide speckle-free illumination when necessary. In schematic (a), the left half of the sample plane shows HC-R speckles used for si-SFDI optical properties, and the right half shows LC-RGB speckle projection for active stereo profilometry. HC-R speckles were contrast-enhanced for visualization. \textbf{(b)} Active stereo predicted surface distances (blue circles) compared to the expected ground truth surface distances (line). \textbf{(c)} Active stereo predicted surface angles for front-tilt (blue circles) and side-tilt (red triangles) compared to the expected theoretical angles (line).}
\end{figure}

We evaluated our 3D profile reconstruction method using a flat homogeneous phantom ($\mu_{a}= 0.013mm^{-1}, \mu^{'}_{s}= 1.26mm^{-1}; 639nm$) that was vertically translated or rotated in both the XZ (front tilt) and YZ (side tilt) planes. In the XZ plane, positive angles indicated the phantom tilting toward the LC-RGB illumination source, while negative angles indicated tilting away, resulting in asymmetric illumination. In the YZ plane, positive and negative angles represent tilts perpendicular to the illumination source, resulting in symmetric illumination. The 3D profile was reconstructed using LC-RGB speckle active stereo images, as shown in Figure~\ref{fig:figure3}(a). We vertically translated the phantom over a $4cm$ range, from $5.8cm$ to $9.8cm$, in $1cm$ increments. As shown in Figure~\ref{fig:figure3}(b), our method reconstructed the distance to within $0.7mm$ of its theoretical value. For side tilt angles ranging from $-40$ degrees to $+40$ degrees, the phantom was reconstructed to within $2$ degrees or less of its theoretical angle, as shown in Figure~\ref{fig:figure3}(c). However, for front tilt angles greater than $+30$ degrees or less than $-30$ degrees, we observed a higher deviation of $3$ to $6$ degrees. To obtain the measured height and angle, we sampled a central $100 \times 100$-pixel ROI.

\subsection{Profile-correction of flat tissue-mimicking phantom} \label{ssec:5.3}

\begin{figure}
\begin{center}
\begin{tabular}{c}
\includegraphics[width=1\linewidth]{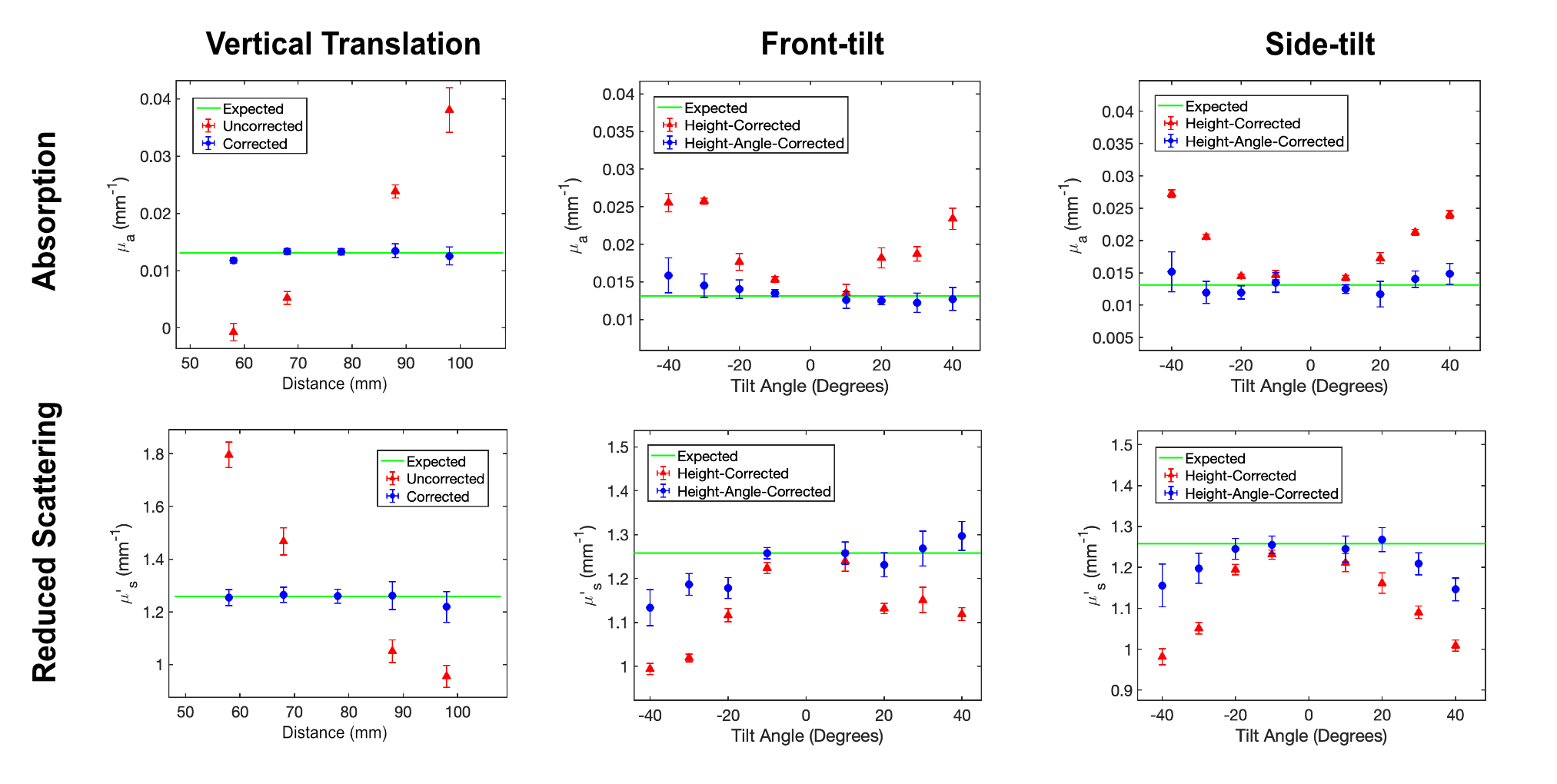}
\end{tabular}
\end{center}
\caption 
{ \label{fig:figure4}\textbf{Profile-correction of optical properties for a vertically translated and angled tissue-like phantom:} Mean and standard deviation of uncorrected (triangles) and profile-corrected (circles) absorption coefficients (top row) and reduced scattering coefficients (bottom row) of a homogenous phantom at 639 nm. The solid line depicts the expected ground truth optical property values measured using conventional SFDI. The left column demonstrates height correction for vertical translations, while the middle and right columns show angular corrections for front tilt and side tilt, respectively.}
\end{figure} 

Without geometrical correction, optical property estimation accuracy decreases when the sample's height or angle differs from the calibration phantom (Fig.~\ref{fig:figure4}). A calibration phantom was measured at distances from $5.8cm$ to $9.8cm$ in $1cm$ increments, all at $0$ degrees of front and side tilt. Height correction was applied using the virtual calibration phantom model described in Section~\ref{ssec:3.2}. The left column of Figure~\ref{fig:figure4} shows the results for uncorrected and height-corrected optical property measurements for a phantom ($\mu_{a} = 0.013mm^{-1}, \mu^{'}_{s}= 1.26mm^{-1}; 639nm$) with different optical properties than the calibration phantom ($\mu_{a}= 0.024mm^{-1}, \mu^{'}_{s}= 0.99mm^{-1}; 639nm$). For uncorrected measurements, as the sample moves closer to the camera relative to the reference phantom's position, absorption is increasingly underestimated, and reduced scattering is increasingly overestimated. The opposite occurs when the sample moves further away from the camera relative to the reference phantom's position. Height-based correction significantly reduced the absorption coefficient's mean error per $cm$ from $78.5\%$ to $2.8\%$ over the entire $4cm$ range. The reduced scattering coefficient mean error per $cm$ decreased from $21.6\%$ to $1.3\%$ over the same range. For absorption coefficients, the mean of the coefficients of variation decreased from $20.9\%$ to $6.5\%$, while for reduced scattering coefficients, it decreased from $4.4\%$ to $3.5\%$. 
\\
\noindent
\\
To test angle-dependent intensity correction, we placed our flat sample on 3D-printed wedges, varying both front and side tilt from $-40$ degrees to $+40$ degrees in $10$-degree increments. We isolated the contribution of angular correction by comparing only height-corrected to height and angle-corrected optical properties. For the remainder of this manuscript, we will refer to the latter as 'profile-corrected' optical properties. In the middle column of Figure~\ref{fig:figure4}, for front tilt from $-40$ degrees to $+40$ degrees, the correction reduced absorption coefficient mean percent errors from $60.5\%$ to $10.4\%$ for negative angles and $40.7\%$ to $4.7\%$ for positive angles. Angle correction also decreased reduced scattering coefficient mean percent errors from $14.3\%$ to $5.5\%$ for negative angles and from $7.8\%$ to $1.5\%$ for positive angles. The positive and negative angle error for the front tilt condition is asymmetric. This asymmetry arises because the HC-R illumination is oriented at a $+15$ degree front tilt from the camera's optical axis (Figure~\ref{fig:figure3}(a)), while our simplified angular correction model assumes coincident camera and source vectors. In the right column of Figure~\ref{fig:figure4}, for side tilts ranging from $-40$ degrees to $+40$ degrees, angle correction reduced the mean percent error of the absorption coefficient from $47.4\%$ to $9.3\%$, and of the reduced scattering coefficient from $11.1\%$ to $3.6\%$. The errors for positive and negative side tilt angles were combined because side tilt rotation is perpendicular to the HC-R illumination source vector, resulting in symmetrical errors. Furthermore, even though the mean accuracy improved, the results show an increase in the mean coefficients of variation when comparing the profile-corrected to the height-corrected results for all experimental groups: front tilt, side tilt, positive angles, and negative angles. For absorption, the increase goes from $4.1\%$ to $10.6\%$, and similarly, for reduced scattering, from $1.5\%$ to $2.5\%$ for only height-corrected to profile-corrected, respectively. The increase in the mean of the coefficients of variation is due to the noise of the estimated surface normals, which propagates into the profile-corrected optical property maps through the angular correction factor.

\subsection{Profile-correction for multiple flat phantoms at different heights} \label{ssec:5.4}

\begin{figure}
\begin{center}
\begin{tabular}{c}
\includegraphics[width=1\linewidth]{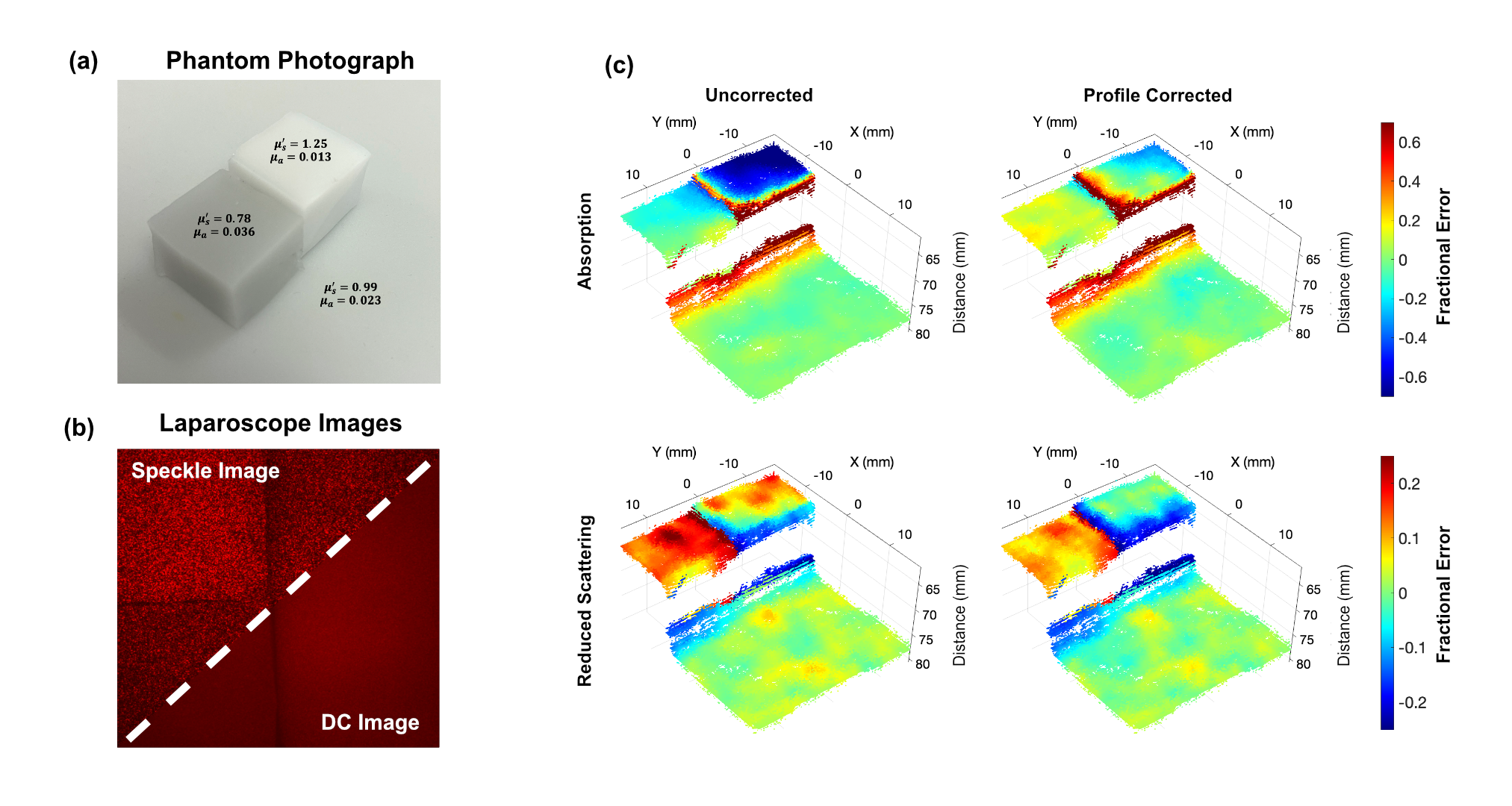}
\end{tabular}
\end{center}
\caption 
{ \label{fig:figure5}\textbf{Multi-height flat phantoms with profile-correction: (a)} Photograph of the three tissue-mimicking phantoms annotated with their optical properties determined by conventional SFDI. \textbf{(b)} Multi-phantom laser speckle (AC; top left) and uniform illumination (DC; bottom right) images used for si-SFDI optical property estimation. The AC data is contrast-enhanced for visualization. \textbf{(c)} Overlay of the optical property fractional error map onto the active stereo estimated 3D point cloud of uncorrected (left column), profile-corrected (right column). Fractional error was computed between laparoscopic si-SFDI and ground truth optical properties calculated from profile-corrected SFDI.}
\end{figure} 

To evaluate profile-corrected laparoscopic si-SFDI in a more complex scenario, we imaged three flat phantoms at different heights within the same FOV. The background phantom was positioned at the reference calibration height (distance = $7.8cm$), while the two other cubic phantoms were placed at smaller distances. Figure~\ref{fig:figure5}(a) shows an oblique photograph of the setup with optical property annotations. As all samples were flat, their surface normals were parallel to the optical axis, resulting in angular correction factors close to unity. Figure~\ref{fig:figure5}(b) shows the speckle and DC image captured by the laparoscope. The speckle image is contrast-enhanced for visualization. Figure~\ref{fig:figure5}(c) shows the fractional error between laparoscopic si-SFDI and ground-truth SFDI optical properties. It shows absorption (top row) and reduced scattering (bottom row), both with (right column) and without (left column) profile-correction. These errors are overlaid on the 3D sample surface as estimated by active stereo. Referring to Figure~\ref{fig:figure5}(c), for the darkest color phantom (Figure~\ref{fig:figure5}(a); $\mu_{a}= 0.036mm^{-1}, \mu^{'}_{s}= 0.78mm^{-1}; 639nm$), profile-correction reduced the fractional error from a $16.1\%$ underestimation of absorption to an overestimation of $10.3\%$, with an overall $5.8\%$ improvement. The reduced scattering error is reduced from an overestimation of $18.7\%$ to $11.2\%$. For the lightest color phantom (Figure~\ref{fig:figure5}(a); $\mu_{a}= 0.013mm^{-1}, \mu^{'}_{s}= 1.25mm^{-1}; 639nm$), we observed a dramatic improvement in absorption from an underestimation of $58.8\%$ to $4.6\%$, while the reduced scattering improved from a $7.4\%$ overestimation to a $3.7\%$ underestimation. Lastly, as a control, the background phantom (Figure~\ref{fig:figure5}(a); $\mu_{a}= 0.023mm^{-1}, \mu^{'}_{s}= 0.99mm^{-1}; 639nm$) shows no significant change in fractional error between the uncorrected and profile-corrected conditions. This is expected because the background phantom is located at the reference calibration height and, therefore, should not undergo a height correction.

\subsection{Hemisphere correction}

\begin{figure}
\begin{center}
\begin{tabular}{c}
\includegraphics[width=1\linewidth]{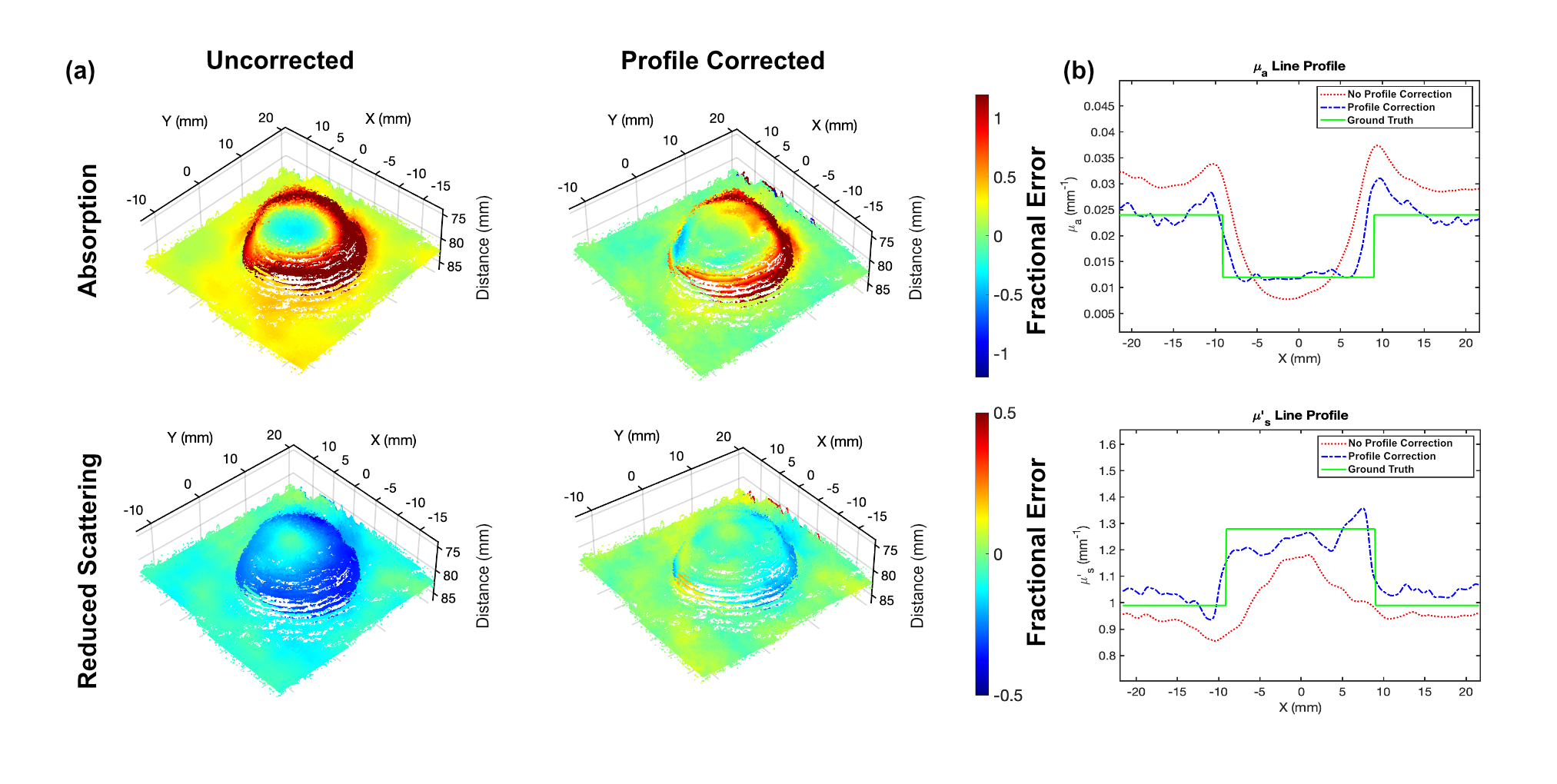}
\end{tabular}
\end{center}
\caption 
{ \label{fig:figure6}\textbf{Hemispheric phantom with profile-correction. (a)}  Overlay of the fractional error map onto the active stereo estimated point cloud of uncorrected (left column), profile-corrected (right column), absorption (top row), and reduced scattering (bottom row) coefficients. \textbf{(b)} Cross-sectional line profile plots through the central region of the hemisphere show the absorption (top row) and reduced scattering (bottom row) coefficients for uncorrected (red dash-dot line), profile-corrected (blue dashed line), and ground truth (green solid line) data.}
\end{figure} 

We tested the profile-correction algorithm on a hemisphere phantom ($\mu_{a}= 0.012mm^{-1}, \mu^{'}_{s}= 1.28mm^{-1}; 639nm$) with both height and surface normal variations (Figure~\ref{fig:figure6}). The top of the hemisphere was placed near the reference calibration height, and this hemisphere was placed on a flat phantom with different optical properties ($\mu_{a}= 0.023mm^{-1}, \mu^{'}_{s}= 0.99mm^{-1}; 639nm$). Figure~\ref{fig:figure6}(a) shows the fractional error maps between laparoscopic si-SFDI and ground truth SFDI optical properties overlaid on the active stereo-estimated surface point cloud. For polar angles less than $40$ degrees, profile-correction reduced the mean percent error for the absorption coefficient from $39.7\%$ to $24.0\%$, and for the reduced scattering coefficient from $16.1\%$ to $4.8\%$. Moreover, profile-correction decreased the coefficient of variation from $53.7\%$ to $30.5\%$ for the absorption coefficient and from $7.9\%$ to $4.1\%$ for the reduced scattering coefficient. Profile-correction fails for absorption at polar angles greater than $40$ degrees, likely due to inaccuracies in steep surface normal estimations and unaccounted object-to-object diffuse reflectance at the intersection of the hemisphere and flat phantom, which are not addressed by our Lambertian reflectance correction model. For polar angles greater than $40$ degrees, the mean percent error after profile-correction remained over $70\%$, necessitating a higher complexity angular correction scheme, which will need to be explored in future studies. In contrast, for polar angles greater than $40$ degrees, this had a less detrimental effect on reduced scattering coefficient, as the profile-correction algorithm reduced the mean percent error from $28.3\%$ to $8.9\%$. The reduced scattering coefficient is less sensitive to steep surface normal angles compared to the absorption coefficient, likely due to the size of the sliding window used for $0.2mm^{-1}$ spatial frequency resolution relative to the profile variations of the phantom. Additionally, the radial averaging of the ACF may smooth out steep angular variations in the local surface structure. It should also be noted that, even after profile-correction, the error for the hemispherical phantom was not radially uniform, with errors being higher in the back. This phenomenon is due to the low signal caused by shadowing when using a single source of angled illumination. The background phantom, on which the hemisphere is placed, also underwent a height correction. This correction reduced the mean percent error for the absorption coefficient from $27.8\%$ to $8.8\%$ and from $8.1\%$ to $3.9\%$ for the reduced scattering coefficient. Figure~\ref{fig:figure6}(b) shows a cross-sectional line profile through the central region of the hemisphere, comparing the ground truth data (green solid line), uncorrected data (red dash-dot line), and profile-corrected data (blue dashed line). The absorption line profile demonstrates how the vertical correction shifts the optical properties towards the ground truth values, while the angular correction transforms the curved profile into a flat, rectangular shape. Similarly, the reduced scattering profile exhibits the same corrective trends, albeit to a lesser extent.

\subsection{\textit{In-vivo} optical property measurements}

\begin{figure}
\begin{center}
\begin{tabular}{c}
\includegraphics[width=1\linewidth]{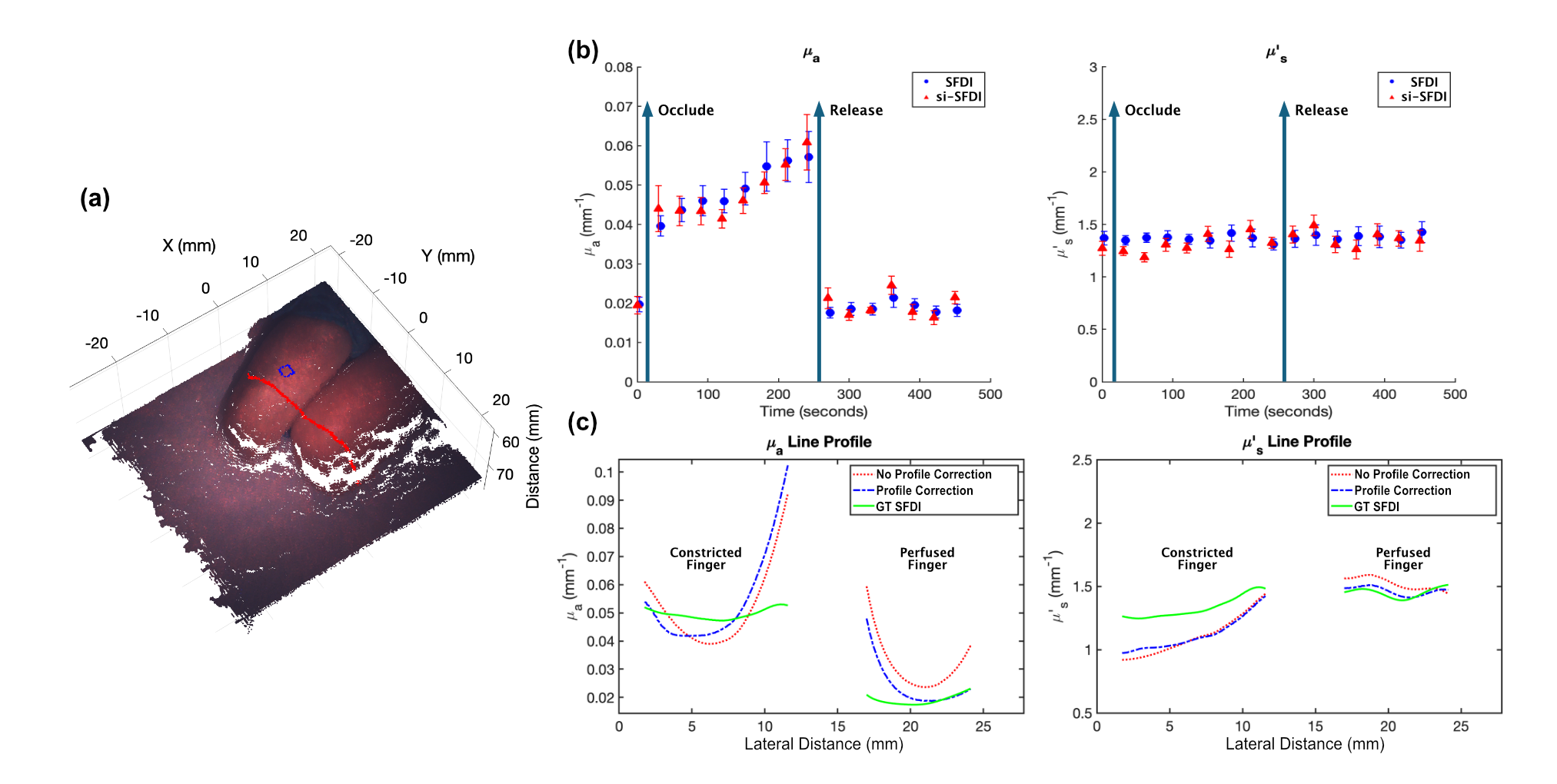}
\end{tabular}
\end{center}
\caption 
{ \label{fig:figure7}\textbf{Profile-corrected, time-series, \textit{in-vivo} finger occlusion study. (a)} Left stereo RGB speckle image overlaid onto the active stereo-reconstructed point cloud. The blue square indicates the ROI analyzed in (b), and the red line represents the line profile for (c) at $t = 180$ seconds. \textbf{(b)} Time-series comparison of profile-corrected laparoscopic si-SFDI and ground truth SFDI estimated absorption (left) and reduced scattering (right) coefficients. Tourniquet constriction began after the $t = 0$ seconds measurement (first arrow) and was removed after the $t = 240$ seconds measurement (second arrow). \textbf{(c)} Comparison of absorption (left) and reduced scattering (right) coefficient profiles at $t = 180$ seconds. Each plot shows uncorrected laparoscopic si-SFDI (red dash-dot line), profile-corrected laparoscopic si-SFDI (blue dashed line), and ground truth profile-corrected SFDI (green solid line). }
\end{figure} 

The profile-corrected optical properties of a complex \textit{in-vivo} object, consisting of two human fingers with the middle finger constricted by a tourniquet, were measured at multiple time points. The decrease in the absorption coefficient at $639nm$ revealed the expected physiological changes due to constriction. Figure~\ref{fig:figure7}(a) shows the left stereo RGB speckle image overlaid onto the active stereo-estimated point cloud of two fingers from the right hand placed on top of a flat, homogenous phantom. In these experiments, the middle finger (top finger) was constricted using a tourniquet, which can be seen in the top corner of Figure~\ref{fig:figure7}(a). At the time point shown in this image, the middle finger had been constricted for $180$ seconds. The ring finger (bottom finger) was perfused for the entire experiment. The blue box shows the $40 \times 40$-pixel ROI used to calculate the mean and standard deviation of optical properties, as shown in Figure~\ref{fig:figure7}(b) at $180$ seconds. The red line shows the line profile location across the constricted middle finger and perfused ring finger used for Figure~\ref{fig:figure7}(c). To assess laparoscopic si-SFDI's sensitivity to temporally changing absorption optical properties due to physiological changes associated with blood flow constriction, Figure~\ref{fig:figure7}(b) presents profile-corrected optical properties of the constricted middle finger measured at $30$-second intervals for $16$ consecutive measurements, spanning a total of $450$ seconds. The optical properties were obtained using both profile-corrected laparoscopic si-SFDI (red triangles) and ground truth from the commercial SFDI system with profile correction (blue circles). The tourniquet was applied after the $t = 0$ seconds measurement (first arrow) and removed after the $t = 240$ seconds measurement (second arrow). The hand was moved between the laparoscopic si-SFDI system and the commercial SFDI system after each measurement, and a $3$ second offset was added to each time stamp to account for the time to switch between systems. Note that the ground truth SFDI markers are offset by $3$ seconds to account for the time delay caused by the slight movement of the hand to the SFDI imaging FOV and the subsequent image acquisition. The mean percent error between laparoscopic si-SFDI and ground truth SFDI was $8.2\%$ and $5.8\%$ for absorption and reduced scattering, respectively. As the finger constriction began, we observed a rapid increase in absorption of $0.025mm^{-1}$ within the first $30$ seconds, followed by a gradual rise of $0.017mm^{-1}$ over the next $210$ seconds (Figure~\ref{fig:figure7}(b). Upon tourniquet release, absorption returned to the baseline value of $0.020mm^{-1}$ within $30$ seconds. These observations at $639nm$ align with known physiological responses associated with increased deoxyhemoglobin and total hemoglobin levels. The gradual absorption increase during constriction and rapid post-occlusion recovery to baseline are consistent with findings reported in Ikawa \textit{et al.}~\cite{ikawa2015relation} In contrast, as expected, the reduced scattering (Figure~\ref{fig:figure7}(b); right plot) did not change during constriction and remained at a baseline value of approximately $1.33mm^{-1}$. To assess the performance of our profile-correction algorithm on \textit{in-vivo} data and examine the spatial changes in optical properties due to constriction, Figure~\ref{fig:figure7}(c) presents cross-sectional line profiles for absorption and reduced scattering. The profiles compare uncorrected data (red dash-dot line), profile-corrected data (blue dashed line), and ground truth profile-corrected SFDI data (green solid line). This figure highlights the changes in absorption at $639nm$ between the constricted and perfused finger at a single time point ($t = 180$ seconds). In the absorption and reduced scattering line profiles (Figure~\ref{fig:figure7}(c)), we thresholded low signals prevalent due to shadowing at the cracks and edges of raised surfaces. The threshold was determined by analyzing the histogram of the flat-field corrected image. For absorption (Figure~\ref{fig:figure7}(c); left plot), we saw slight improvements due to height-correction, but minimal changes from angular-correction. In contrast, for reduced scattering (Figure~\ref{fig:figure7}(c); right plot), the profile-correction algorithm did not yield significant improvements between the uncorrected and profile-corrected line profiles. To obtain overlaid images for consistent line profile analysis, ground truth SFDI optical property maps were registered to si-SFDI optical property maps through an affine transformation. The line profile represents an average from a rectangle with a $10$-pixel thickness.

\section{Discussion}
\label{sect:discussion}
In this work, we introduced a novel stereo laparoscope implementation of 3D profile-corrected laser speckle si-SFDI. The proposed approach enables the simultaneous measurement of tissue topography and profile-corrected optical properties through a commercial stereo laparoscope. This method employed a coherent laser coupled into a multimode fiber to project an objective speckle pattern onto the sample. We characterized the tissue response in the spatial frequency domain by analyzing the PSD of each speckle image, which allows tissue optical properties to be calculated through phantom calibration \cite{chen2021speckle}. For non-flat samples, we used active stereo to estimate the sample's 3D profile and then used this profile to correct for height- and angle-dependent reflectance variations.
\\
\noindent
\\
Our experiments showcased the initial critical steps necessary for the clinical implementation of si-SFDI as a non-invasive, quantitative contrast imaging technology for surgical guidance. First, we validated laparoscopic si-SFDI against ground truth SFDI on $13$ homogeneous tissue-mimicking phantoms and observed an average error of $6.4\%$ for absorption and $5.8\%$ for reduced scattering. This validation across diverse optical properties demonstrated our method's accuracy, which is crucial for surgical guidance, as the precision is sufficient to differentiate disease states like carcinoma \cite{rohrbach2014preoperative}. Second, during surgery, the sample’s distance is typically heterogeneous and changes continuously due to manipulation of the surgical field. Uncorrected results in Figure~\ref{fig:figure4} demonstrated that optical property estimation for samples at distances different from the calibration phantom requires profile-correction. As the sample distance deviated from the reference calibration distance, the absorption and reduced scattering mean error per $cm$ reached $78.5\%$ and $21.6\%$, respectively. Applying the height correction over a $4cm$ range reduced the mean error per $cm$ to $2.8\%$ for absorption and $1.3\%$ for reduced scattering coefficients. Third, to assist the surgeon, laparoscopic si-SFDI should be able to distinguish different tissue types at various heights based on their optical properties within a single FOV. To validate this capability, we tested our method on complex phantom arrangements with varying optical properties within a single FOV: multiple flat phantoms at different heights (Figure~\ref{fig:figure5}) and a hemispheric phantom (Figure~\ref{fig:figure6}). The profile-correction algorithm maintained the mean absolute error within $0.007mm^{-1}$ for absorption and $0.11mm^{-1}$ for reduced scattering maps. Lastly, monitoring wide-field optical properties, especially absorption, over time is essential for confidently assessing tissue viability and detecting early-onset ischemia \cite{gioux2011first, pharaon2010early, yafi2011postoperative}. Figure~\ref{fig:figure7} demonstrates the ability of laparoscopic si-SFDI to accurately monitor temporal changes in tissue absorption resulting from hypoxia induced by vessel constriction. Moreover, unlike local contact techniques such as tissue oximetry, laparoscopic si-SFDI can identify ischemic regions over a wide FOV without prior knowledge of the ischemic region's location, making it attractive for image-guided surgery.
\\
\noindent
\\
The results are promising and have demonstrated feasibility in phantom and simple \textit{in-vivo} studies; however, many factors should be considered to further strengthen this technique. First, to estimate the 3D profile for downstream profile-correction, we restricted our maximum estimated angle to $40$ degrees (Figure~\ref{fig:figure3}). At tilt angles of $\pm 30$ degrees and $\pm 40$ degrees (Figure~\ref{fig:figure3}(c)), our camera's limited dynamic range struggled to capture both nearby and distant details of the tilted phantom simultaneously. Therefore, the active stereo-reconstructed point clouds exhibited high levels of noise and error, which propagated through the optical property correction algorithm. An important tradeoff during the profile estimation is the resolution versus accuracy of the measurement. To reduce noise, we downsampled our point cloud through smoothing and grid-fitting \cite{jderrico2016}. However, this caused the profile estimations to lose important surface normal information at steep steps and curved object interfaces. This was evident in the hemisphere phantom (Figure~\ref{fig:figure6}), the crack between the constricted and perfused fingers (Figure~\ref{fig:figure7}(a/c)), and other areas with significant height changes (Figure~\ref{fig:figure5}). This tradeoff between resolution and accuracy can be mitigated by imaging at smaller distances, though this approach reduces the FOV. Second, even with correct angular surface normal estimation, there were errors associated with angled, single-source illumination (Figure~\ref{fig:figure4}) and inter-object diffuse reflectance (Figure~\ref{fig:figure6}), which exposed the limitations of the Lambertian reflectance assumptions governing our profile-correction algorithm. In Figure~\ref{fig:figure4}, we observed asymmetrical errors in optical property estimations with sample front tilt. These asymmetries could be circumvented by illuminating parallel to the detection axis by utilizing the commercial fiber light guide that is parallel and close to the laparoscope objective lenses. This solution may also reduce barriers with pilot clinical studies since it may be possible to project laser speckle patterns through the laparoscope fiber guide, leaving the distal instrument unmodified. In Figure~\ref{fig:figure6}, we observed large errors in the profile-corrected absorption at the junction of the hemisphere and flat phantom, due to the algorithm's insensitivity to optical property changes caused by inter-object diffuse reflectance. Future work could integrate a modified Lambertian correction \cite{zhao2016angle}, which demonstrated accurate angular correction up to $75$ degrees in SFDI by adding a correction factor to the Lambertian correction term, accounting for inter-object diffuse reflectance. Third, our results demonstrated sensitivity to optical property changes when phantoms with diverse optical properties were placed at multiple heights within the same FOV (Figure~\ref{fig:figure5}). This sensitivity was observed when the phantoms' optical properties were chosen within a $50\%$ absorption and $25\%$ reduced scattering difference of the calibration phantom. While these results highlight the sensitivity of our profile-corrected si-SFDI method for optical property combinations within this range, we did not test the range at which the error would significantly increase. This limitation is primarily due to the dynamic range of the $8$-bit (per channel) color camera used in this proof-of-concept imaging system. To address this challenge and enhance the system's performance, future studies could employ a 16-bit monochrome camera, which will not only resolve the dynamic range issues but also double the sampling resolution of our laser speckle, enabling the use of smaller objective speckle grains for better local optical property sensitivity.
\\
\noindent
\\
There are several areas for improvement of laparoscopic si-SFDI in order to make it amenable to clinical testing. First, it is important to have video-rate data acquisition. The current prototype requires the laser speckle reducer to be toggled on and off to sequentially acquire an AC and DC image, which takes about $1$ second. A spinning half-diffuser could be used to increase acquisition speed, allowing for optical property data acquisition at half of the rotational frequency of the optical chopper. Alternatively, it may be possible to estimate optical properties from a single speckle image-–previous work \cite{bobrow2019deeplsr} has shown that a DC-illumination can be estimated from a speckle image using a content-aware deep learning approach. Second, it is important to have faster data processing. The current processing speed for profile-corrected si-SFDI is slow due to stereo matching, 3D reconstruction, and sequential window-based computations. It takes about $12$ seconds to estimate $784 \times 960$-pixel profile-corrected optical property maps from raw data acquisitions using a $4$-core $3.6$ GHz processor in MATLAB. However, since each window computation is independent, this algorithm is compatible with parallelization using a graphics processing unit. Another possibility is to explore an end-to-end deep learning model trained to directly predict profile-corrected optical properties from a single speckle image. Earlier work by Chen \textit{et al.}. \cite{chen2019ganpop} demonstrated that a single structured light image could be used to predict optical properties in real-time. Furthermore, Aguénounon \textit{et al.} \cite{aguenounon2020real} showed that a twin network could perform both surface profile correction and accurate optical property extraction from a single structured light image. Third, it is necessary to optimize the objective speckle contrast to enhance the sensitivity and accuracy of si-SFDI optical property estimation for diverse samples within the same FOV. To enhance objective speckle contrast, we could potentially use beam shaping techniques to preferentially excite higher-order modes while attenuating lower-order modes\cite{florentin2017shaping}.
\\
\noindent
\\
Clinical adoption of our technique as a noninvasive quantitative surgical guidance tool relies on advancing it from laparoscopic optical property estimation to providing clinically relevant surgical information, such as wide-field oximetry measurements and quantitative margin assessments in fluorescence-guided surgery. Incorporating a second wavelength would enable tissue oximetry imaging, which was originally demonstrated in SFDI \cite{gioux2011first}. Furthermore, when coupled with fluorescence-based contrast mechanisms, optical properties can be used to correct emitted fluorescence, providing quantitative fluorescence estimates\cite{valdes2017qf} and 3-D localization of fluorescently labeled subsurface tumors and structures\cite{o2022quantitative, petusseau2024subsurface}.

\section{Conclusion}
\label{sect:conclusion}
In this work, we presented a speckle-illumination stereo laparoscope that acquires profile-corrected optical properties. We detailed the system's design principles, data acquisition, and processing workflow, and validated its accuracy on tissue-mimicking phantoms with complex geometries against a commercial SFDI system as ground truth. Furthermore, we demonstrated the system's capability for \textit{in-vivo} imaging of time-varying physiological changes. By building upon the principles of speckle-illumination wide-field imaging of optical properties, this work lays the foundation for image-guided minimally invasive surgery using quantitative endogenous contrast.

\section{Disclosures}
The authors declare that there are no conflicts of interest related to this article.

\section{Code, Data, and Materials Availability}
The data supporting this article’s findings are available upon reasonable request from the corresponding author.  

\section{Funding}
This work was partially supported with funding from the NSF CAREER Award 2146333, the NSF Graduate Research Fellowship Program (DGE-1746891), and a Technology Research Grant provided by Intuitive Surgical.


\section{References} \label{sect:references}

\bibliography{report}   
\bibliographystyle{spiejour}   

\section{Biographies}

\vspace{2ex}\noindent\textbf{Anthony A. Song} is a 3rd year PhD student in the Biomedical Engineering Department at The Johns Hopkins University. His research interests include biophotonics and machine learning.

\vspace{2ex}\noindent\textbf{Mason T. Chen} is an Applied Scientist at Amazon. He obtained his PhD in Biomedical Engineering from Johns Hopkins School of Medicine in 2022.

\vspace{2ex}\noindent\textbf{Taylor L. Bobrow} completed his PhD in biomedical engineering at the Johns Hopkins University. He is currently a postdoctoral research fellow in the Computational Biophotonics Lab at the Johns Hopkins University. His research interests include biophotonics, computer vision, computational imaging, and artificial intelligence.

\vspace{2ex}\noindent\textbf{Nicholas J. Durr} is an Associate Professor of Biomedical Engineering at Johns Hopkins University. He received his BS degree in electrical engineering and computer science from UC Berkeley in 2003 and MS and PhD degrees in biomedical engineering from UT Austin in 2007 and 2010, respectively. His research interests include medical devices, machine learning, and biophotonics.

\end{spacing}
\end{document}